# Comment on « Gain coefficient method for amplified spontaneous emission in thin waveguided film of a conjugated polymer » [APL 93, 163307 (2008)]

Sébastien Chénais and Sébastien Forget

*Laboratoire de Physique des Lasers, UMR 7538 du CNRS, Institut Galilée, Université Paris-Nord, 99 avenue Jean-Baptiste Clément, 93430 Villetaneuse, France.*

In a recent letter, Silvestre *et al.* address the interesting issue of finding a general technique to fit the experimental data obtained with the Variable Stripe Length technique, which is commonly used to measure gains in waveguide lasers [1]. Relying on the one-dimensional equation

$$\frac{dI_{out}}{dz} = \alpha I + g(z) I_{out} \qquad (1)$$

($I_{out}$ is the edge emitted intensity; $I$ is the incident pump intensity; $g$ is the net gain; α is related to spontaneous emission), the authors claim that the gain $g$ appearing in equation (1) depends on the excitation stripe length.

The letter contains several errors, misconceptions and confusing statements.

Firstly, it is obvious that the gain, as defined in eq.(1), cannot depend on the *length* of the stripe, but only on the position within the stripe. The net gain $g$ appearing in eq (1) is fundamentally a local parameter, defined by $g(\vec{r}) = \sigma_{em} \Delta N(\vec{r}) - \eta$, where $\sigma_{em}$ is the emission cross section, $\Delta N$ the local population inversion density (in $cm^{-3}$), and η the losses ($cm^{-1}$). When the 1-dimensional approximation is valid, g is actually only dependant on the position z. As a consequence, in a transversely-pumped thin slab geometry used by the authors, the gain — averaged over the waveguide thickness — is related to the pump intensity (in W/cm²) and not to the total pump power, a misconception that leads to a supposed expression for the gain written as $g(z) = 2b(z).z.w.I$ where z is the stripe length and w the stripe width. The introduction of b(z) is useless since g(z) has no particular reason to be linear with z. In contrast, the inhomogeneous pump profile (in general a truncated Gaussian along the stripe direction, with possible diffraction effects added by the slit) should be taken into account if the pump intensity cannot be made constant, as well as gain saturation (upon introduction of a saturation intensity), which may affect the shape of g(z). Examples of such analyses may be found in refs [2, 3].

Secondly, the reason why the term $\alpha I$ of equation (1), accounting for spontaneous emission (SE), is replaced by an Arrhenius expression of the form $\alpha I_{max} / (1 + Ce^{-I/I_0})$ is not clearly justified, and is not physically reasonable. This expression involves that when the pump laser is off (*I = 0*), the emitted intensity should grow linearly, according to eq. (1), which has obviously no physical meaning. It is exact that spontaneous emission may not scale linearly any more with *I*, for instance when stimulated emission is strong enough to significantly deplete the excited state population [4]. This gain-saturation related effect is in general neglected since it corresponds to a regime where

SE is negligible itself compared to stimulated emission. Whatever the new form proposed for this term, the linear dependence with the pump intensity *I* should be preserved at low *I* values.

Finally, the authors introduce four new independent unknown parameters ($\alpha I_{max}$, C, b and $I_0$) and obtain good fits, which is not very surprising. Besides the fact that $I_0$ is an ill-defined parameter, obtained from the Arrhenius expression, it is referred to as a "threshold intensity" (expressed in W/cm²) establishing the crossing from the SE to the Amplified SE regime. It is then, like the gain, a local parameter which cannot depend on the stripe length, as shown in figure 4.

At last, the authors use the 1-dimensional approximation (eq. 1) in limit cases where it is highly questionable. Since the width w of the stripe is given to be 0.02 cm, data obtained for a length L = 0.03 cm cannot be processed with a simple 1-D model without care.

In conclusion, the authors report the experimental evidence of an apparent variation of gain with the stripe length, but do not provide a clear framework for analysis. A thorough analysis of the Variable-Stripe Length technique for the study of ASE in thin films may be found in references [1, 3, 5].